# Scanning micro X-ray diffraction unveils the distribution of oxygen chains nano-puddles in $YBa_2Cu_3O_{6.33}$

Gaetano Campi[1], Alessandro Ricci[2], Nicola Poccia[3], Luisa Barba[4], Gianmichele Arrighetti[4], Manfred Burghammer[5], Alessandra Stella Caporale[6], Antonio Bianconi[6]

[1] *Institute of Crystallography, CNR, via Salaria Km 29.300, Monterotondo Roma, 00015, Italy*
[2] *Deutsches Elektronen-Synchrotron DESY, Notkestraße 85, D-22607 Hamburg, Germany.*
[3] *MESA+ Institute for Nanotechnology, University of Twente, P. O. Box 217, 7500AE Enschede, Netherlands.*
[4] *Elettra Sincrotrone Trieste. Strada Statale 14 - km 163, 5, AREA Science Park, 34149 Basovizza, Trieste, Italy*
[5] *European Synchrotron Radiation Facility, B. P. 220, F-38043 Grenoble Cedex, France.*
[6] *RICMASS Rome International Center for Materials Science Superstripes, via dei Sabelli 119A, 00185 Roma, Italy*

Oxygen chain fragments are known to appear at the insulator-to-superconductor transition (SIT) in $YBa_2Cu_3O_{6+y}$. However the self organization and the size distribution of oxygen chain fragments is not known. Here we contribute to fill this gap, using scanning micro-X-ray diffraction which is a novel imaging method based on advances in focusing synchrotron radiation beam. This novel approach allows us to probe both real-space and k-space of a high-quality $YBa_2Cu_3O_{6.33}$ single crystals with $T_c$=7K. We report compelling evidence for nanoscale striped puddles, with Ortho-II structure, made of chain fragments in the basal Cu(1) plane with local oxygen concentration y≥0.5. The size of the Ortho-II puddles spans a range between 2 and 9 nanometers. The real space imaging of Ortho-II puddles granular network shows that superconductivity, at low hole-doping regime, occurs in a network of nanoscale oxygen ordered patches, interspersed with oxygen depleted regions. The manipulation by thermal treatments of the striped Ortho-II puddles has been investigated focusing on the "spontaneous symmetry breaking" near the order-to-disorder phase transition at $T_0$=350 K.

## 1. Introduction.

High temperature superconductivity (HTS) appears in composite materials made of stacks of active superconducting layers intercalated by spacer layers [1]. In $YBa_2Cu_3O_{6+y}$ (YBCO) [2] the active layers are made of an infinite layer cuprate, $Y[Cu(2)O_2]_2$, intercalated with the "spacer oxide block" made of a defective rocksalt oxide, $[BaO]_2Cu(1)Oi_y$. In the defective "spacer oxide block" the oxygen defect ions $Oi_y$ partially fill the empty basal plane sites. Inhomogeneity is a generic feature of HTS because of the electronic phase separation near the charge transfer Mott phase, where metallic stripes [3], hosting the doped holes with $O(2p^5)Cu(3d^9)$ configuration [4], coexist with antiferromagnetic domains. Moreover, also the lattice diverges from the average structure [5] because of i) unscreened defects in the spacer layers; ii) corrugated $CuO_2$ short range nano-domains driven by polaronic local lattice distortions [6] and the lattice misfit strain between active and spacer blocks [7,8]. While in the past years lattice inhomogeneity was not considered an essential parameter in the search of the mechanism of HTS, recently the interest on the role of defects for controlling the critical temperature is growing [9]. While disorder usually suppresses the superconducting critical temperature ($T_c$) in conventional superconductors, an optimum lattice inhomogeneity has been found to enhance the $T_c$ in HTS [10,11]. The tendency toward electronic phase separation in strongly correlated multiband systems with polaronic and Fermi particles near a Mott transition is now well accepted [12-15]. Recently several theories have been proposed, based on the presence in HTS of networks of nanoscale superconducting grains [16-21].

The nanoscale phase separation related to the self-organization of defects is not a unique feature of cuprates like super-oxygenated $La_2CuO_{4+y}$ [22], $La_{2-x}Sr_xCuO_4$ [6,23,24] and $Bi_2Sr_2CaCu_2O_{8+y}$ [25], but it has been found also in $Al_{1-x}Mg_xB_2$ [26], doped iron-chalcogenides [27-29], and other functional materials like manganites [30]. Thermal treatments control defect organization [31-34] on photo-induced effects [35], muon-spin resonance (μSR) [36] showed that lattice complexity controls $T_c$ in YBCO. However, direct information concerning lattice inhomogeneity in YBCO [37] is missing because of lack of suitable experimental methods.

In this work we have used the novel scanning micro X-ray diffraction (μXRD) [11,22] to unveil the lattice complexity in YBCO. We focus on the size distribution and imaging the spatial distribution of oxygen Oi chains fragments in the basal plane near the insulator-to-superconductor transition (SIT) and we study the effect of thermal treatments [38, 39]. The satellite superstructure streaks in the XRD pattern of YBCO probing the metamorphic phase with local oxygen concentration y=0.5 and called Ortho-II by Nakazawa et al. [40], have been observed in the range 0.33<y<0.62 [41-43]. We have analyzed these superstructures to get the size distribution of the nano-scale Ortho-II puddles and their spatial distribution.

## 2. Materials and Methods.

The single crystals of $YBa_2Cu_3O_{6.33}$ were grown using barium zirconate crucibles by the self-flux technique using chemicals of 99.999% purity for CuO and $Y_2O_3$, and 99.997% for $BaCO_3$ [44,45]. The purity of the crystals was found to be better than 99.99 atomic % by analysis using inductively-coupled plasma mass spectroscopy.

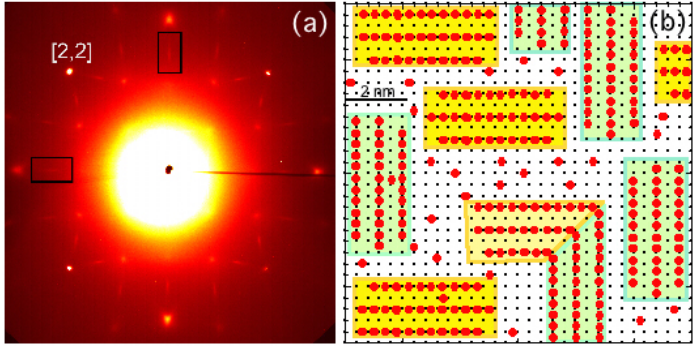

**Figure 1.** (Colored online) Panel (a): the diffraction pattern in the (h,k) plane of $YBa_2Cu_3O_{6.33}$ ($T_c$=7K) measured in transmission mode on a 100 micron thick sample with c-axis oriented in the x-ray beam direction. 20 KeV photon energy and 100 micron x-ray beam size have been used. The diffuse streaks $q_{ortho-II}$ at $(h \pm 0.5, 0)$ are superstructure satellites due to nanoscale puddles with Ortho-II structure made by Oi ordering after annealing at 300 K. The equal intensity of horizontal and vertical streaks highlighted with the black rectangles, indicate a phase made of the Ortho-II puddles where the a-axis is oriented both in the vertical and horizontal direction. Panel (b): pictorial view of the basal Cu-$Oi_y$ plane of YBCO. The small black dots are the Cu(1) sites and the large red dots are the oxygen defect ions $Oi_y$. Horizontal and vertical oxygen rich metamorphic phase (y=0.5) called Ortho-II by Nakazawa et al. [40]. The Ortho-II puddles are intercalated by disordered oxygen poor (y<0.33) domains. The black bar indicates 2 nm length scale.

The oxygen content of the crystal was set to 6.33 by annealing in flowing oxygen at 914°C followed by quenching to room temperature under flowing nitrogen gas. The macroscopic oxygen content inhomogeneity was removed by annealing the crystal at 570° C in a tiny sealed quartz capsule followed by quenching to ice-water bath. The crystal was then kept at room temperature to let the short range oxygen order to develop [46].
We have investigated the sample superstructure due to nanoscale puddles of self organized oxygen ions filling about 1/3 of the sites on the basal plane.
The structure has been investigated with two advanced methods made possible by the use of synchrotron radiation: 1) X-ray diffraction in transmission-mode using a high energy photon beam and 2) scanning micro- X-ray diffraction in reflection.
The source of the 12.4 KeV X-rays for scanning µXRD at the ID13 beamline of ESRF is an electron undulator. The crystal optics includes a tapered glass capillary. It produces a 1 µm beam spot at the sample. A charge-coupled area detector (CCD) records the X-rays scattered by the sample [22].
The source of the 20 KeV X-ray beam was the XRD1 beamline of the Elettra synchrotron in Trieste (Italy). The diffraction patterns as a function of temperature were collected by means of a K-diffractometer with a motorized goniometric X-Y stage head and a Mar-Research 165 mm Charge Coupled Device (CCD) camera, 70 mm far from the sample. The 20 KeV X-ray beam was selected from the source by a double-crystal Si(111) monochromator.[38]

## 3. Results and Discussion

A typical X-ray synchrotron diffraction (XRD) pattern of the $YBa_2Cu_3O_{6.33}$ in transmission mode, probing the bulk structure, is depicted in Fig. 1a. The lattice parameters are: a = 3.851(4) Å, b = 3.856(4) Å, c = 11.78(5) Å. The Cu-O bond distance is 192.8 pm, showing a 2% compression relative with the Cu-O equilibrium distance of 197 pm [47].
Therefore this sample is near the optimum misfit-strain [7,8] and the SIT transition. Visual inspection of the 2D diffraction pattern in Fig.1a shows immediately the satellite reflections $q_{ortho-II}$ (h±0.5,k,l) between the principal reflections associated with ordered oxygen chains fragments forming nanoscale Ortho-II puddles.
These $q_{ortho-II}$ satellites have a needle shape clearly pointing in both horizontal and vertical directions. The equal intensity of vertical and horizontal streaks indicates a phase with an equal number of horizontal and vertical puddles. A pictorial view of the phase separation derived from these data is shown in Fig. 1b. The oxygen rich Oi chain fragments form nanoscale "Ortho-II" puddles coexisting with oxygen poor domains with disordered $Oi_y$.

The point to point spatial variation of the size of the "Ortho-II" puddles has been measured by scanning µXRD in reflection using the 12.4 KeV X-rays focused on a micron size spot and probing a thickness of about 1 micron [22]. The (0.5,0,3) superstructure satellite reflection was measured at each point (x,y) of the sample reached by the x-y translator with micron resolution. The integrated intensity of the superstructure peaks and their position in different spots of the crystal are quite homogenous ($q_{ortho-II}$ =0.5 with a standard deviation of 0.001). This indicates that the nano-scale oxygen puddles have the Ortho-II periodicity and confirms the high quality of the crystal. On the contrary, the full width half maximum (FWHM) of the peaks showed significant variations from point to point: indeed, the domain size of the Ortho-II puddles, derived from the measured FWHM via standard methods of diffraction [43] was found ranging between 3 and 9 nanometers. Upon inspection of the real space mapping of the nanoscale Ortho-II puddles sizes shown in Fig. 2, the heterogeneous granular structure appears quite clearly. The puddle sizes have been derived from the inverse of the superstructure satellites FWHM in the k-space along the a* (panel a), c* (panel b) directions. More specifically, the FWHM of the diffraction profiles in the h-direction probes the size of the Ortho-II puddles transverse to the chains direction i.e. provides the number of chain fragments in a puddle.

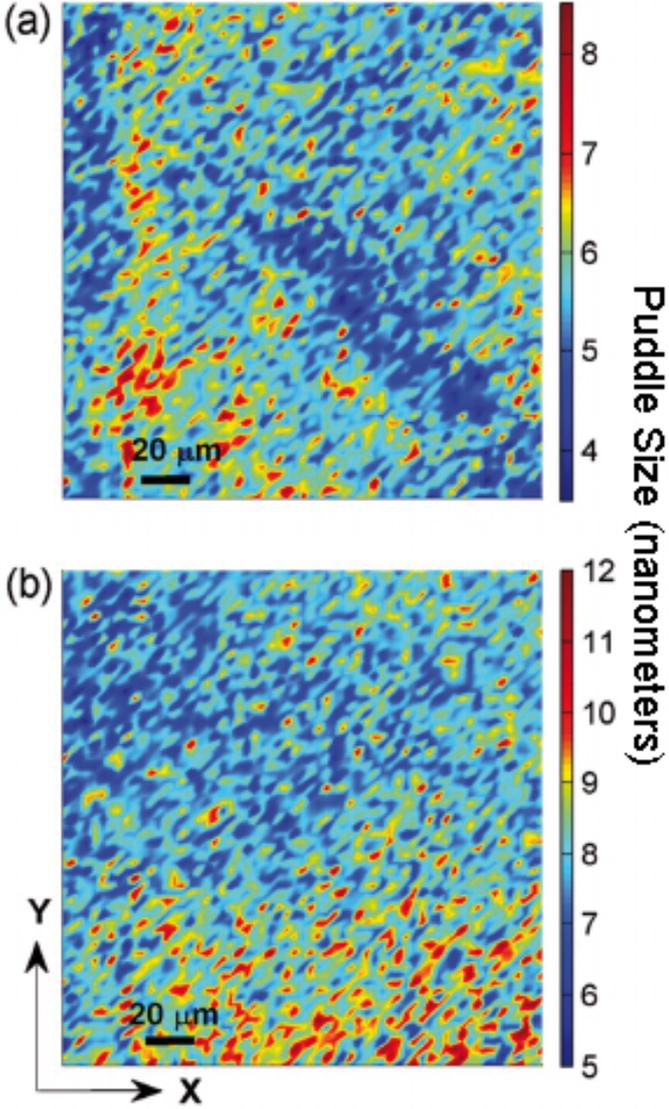

**Figure 2.** Scanning micro X-ray diffraction of a single crystal $YBa_2Cu_3O_{6.33}$ using a 12 KeV energy and probing 1 micron thickness of the crystal surface. The puddle size (color) of the Ortho-II puddles is plotted as a function of the illuminated spot position XY in the sample surface, where x and y direction in the image correspond to the a and b crystallographic axis of the sample. The size of the Ortho-II oxygen chain puddles, averaged over 1 micron illuminated spots, changes from site to site. It ranges from 3 (dark blue) to 12 (red) nanometers in the a-axis direction (panel a) and in the c-axis direction (panel b).. The white bar indicates 20 µm length scale on the sample surface.

At the same time the FWHM in the c*-direction of the reciprocal lattice probes the chain fragment length along the c-axis direction. The statistical analysis of the Ortho-II puddles sizes is shown in Fig. 3: the probability distributions of the length of the chain fragments (Fig. 3a) and their size along the c-axis (Fig. 3b) show that the Ortho-II puddles are made of 2.5 to 12 oxygen chains oriented along the a-axis (Fig. 3a) in domains with average size of 4 nm and a standard deviation of 0.6 nm along the c crystallographic direction (Fig. 3b). We observe clear irregularities and deviations from normal behavior in the distribution tails. In order to quantify these deviation we have calculated the distribution skewness, $sk$, giving $sk_a$=0.3 and $sk_c$=0.9 for the Ortho-II domain size along the a and c directions, respectively. Their positive values indicate a larger weight of the right tails in both distributions, while the fact that $sk_c$ is substantially larger than $sk_a$ confirms accentuated disorder in the out of plane direction.

The size distribution is averaged over a 1 micron area that is much smaller than in any other x-ray or neutron diffraction experiment reported before but it is much larger than the size of the puddles. Therefore, deeper insight could be carried out by investigation with smaller nanometer beams.

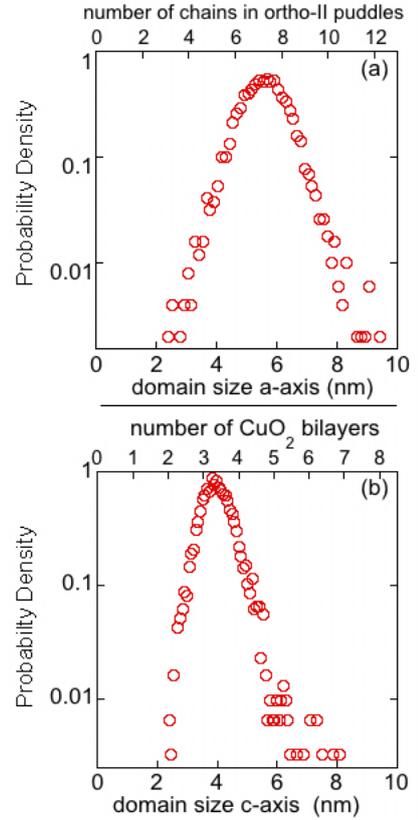

**Figure 3.** Statistical distribution of the size of the Ortho-II puddles from data shown in Fig. 2. Panels (a) and (b) show the probability distribution function of spots in the crystal where the Ortho-II puddles have the same size in the a-axis (c-axis in panel b) direction. The puddle size in nanometers is determined from the FWHM of superstructure diffraction $q_{ortho-II}$ satellite reflections in the a* (c* in panel b) direction. The measured size of the Ortho-II puddles shows that the number of chains is bigger than the number of $CuO_2$ bilayers in each puddle. This is related with a larger mobility of oxygen defects in the basal plane.

Having determined the spatial imaging and distributions of the Ortho-II puddles we move to the effects of thermal treatments. The diffraction patterns as a function of temperature were collected at the XRD1 beamline of the ELETTRA synchrotron in Trieste (Italy). The temperature of the crystals has been increased from 270K to 400 K, using a cryo-cooler (700 series Oxford Cryosystems). The heating cycle was carried out through a quite slow rate (0.2 K /minute). For the temperature cycles the sample was sealed in a oxygen filled ampoule and it was kept only few minutes above 350K in the thermal cycle with any loss of oxygen out of the sample surface and no change in the overall stoichiometry as shown by the fact that the sample back to same phase as the as grown crystal, after a month. Figure 4 shows the three dimensional temperature evolution of the (2.5,0,0) Ortho-II satellite profile along h and k, respectively.

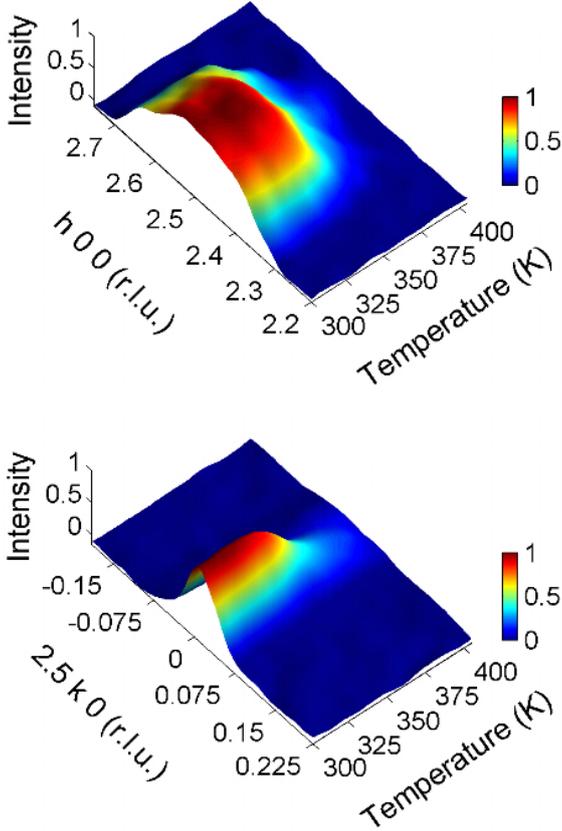

**Figure 4.** Temperature evolution of the $q_{ortho-II}$ satellite reflection profiles along h (upper panel) and k (lower panel). The bright (red-online) and dark (blu on-line) color corresponds respectively to the higher and lower intensities of the satellite diffraction spots. Variation of the average structure of ortho-II puddles in the (h,k) plane of $YBa_2Cu_3O_{6.33}$ measured in transmission mode of a 100 micron thick sample using a X-ray beam of 20 KeV photon energy and 100 micron size.

Although the order to disorder phase transition at $T_0=350\pm5K$, we can appreciate the presence of smaller nanoscale Ortho-II puddles above $T_0$, up to 400K in agreement with previous works [25,26]. After the heating, we completed our thermal cycle with a cooling ramp performed at the same heating rate. The variations of the intensity and of the FWHM of the Ortho-II superstructure diffraction profiles during the whole cycle are plotted in Fig. 5. Here the FWHM in the k-direction of the reciprocal lattice probes the chain fragment length along the b-axis direction. We observed a relevant hysteresis (Fig. 5) demonstrating that once the Ortho-II nano-puddles of oxygen chains are formed, they cannot be reformed in a reversible way, with the same size, in the time scale of the experiment.

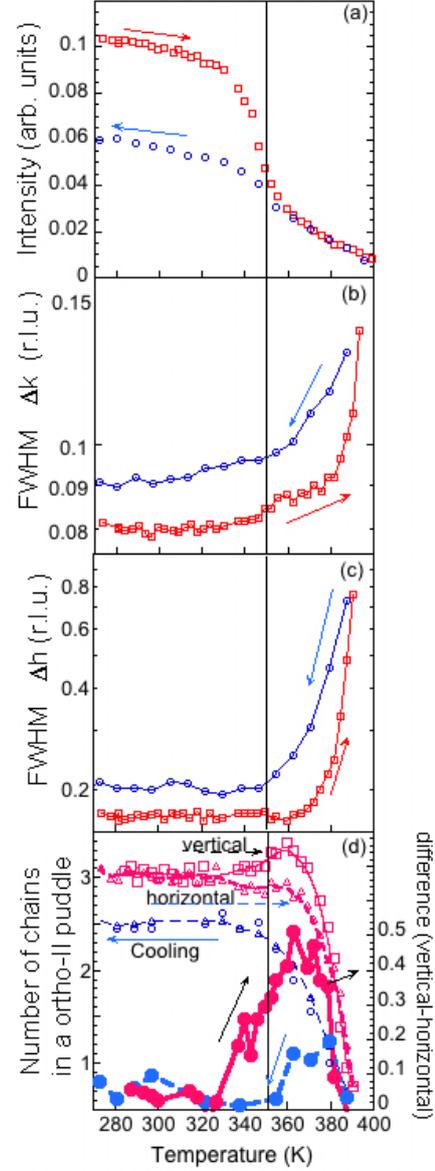

**Figure 5.** Temperature evolution of the $q_{ortho-II}$ superstructure in the heating cycle followed by the cooling cycle. The average normalized intensity as a function of temperature is shown in panel (a). The squares (red online) and circles (blue online) represent the data collected during the heating and cooling thermal cycles, respectively. The average FWHM along b* and a* direction are plotted in panel (b) and (c) respectively. Panel (d) shows the variation of the size of the Ortho-II puddles oriented in the vertical (open squares red-on-line) and horizontal (open triangles) direction in the sample, during the warming cycle. The difference between the size of vertical and horizontal ortho-II puddles indicated by large filled (red-on-line) dots (filled dots, blue on-line) in the heating (cooling) cycle shows a large value in the range 350K<T<380 K due to the spontaneous symmetry breaking in the fluctuation regime above 350 K.

As it has been shown in Fig. 1 the Ortho-II puddles are equally oriented in the horizontal and vertical direction of the sample with y=0.33 giving diffraction streaks broad and diffuse in the h-direction and narrow in the k-direction. The broad and diffuse streaks are observed both in the vertical and horizontal direction, indicating vertical and horizontal puddles. In the as grown sample we observe the same diffraction intensity and the same width of horizontal and vertical streaks showing that the average size of Ortho-II puddles is the same for vertical and horizontal puddles. Increasing the temperature above the order-to-disorder transition the size of the Ortho-II puddles decreases as shown in Fig. 5d. In the heating cycle a large difference between the size of vertical and horizontal Ortho-II puddles appears in the temperature range of 350 K<T<380 K above the order-to-disorder phase transition temperature ($T_0$=350 K). A smaller variation in the 320 K<T<350 K temperature range, below the phase transition temperature, is observed (Fig. 5d). This large anisotropy due to preferential ordering emerges in the fluctuation regime above the order-to-disorder critical temperature $T_0$=350K. This phenomenon provides evidence for "spontaneous symmetry breaking" in the fluctuation regime near the order-to-disorder phase transition temperature.

## 4. Discussion.

We have investigated the variation of the Ortho-II puddles size from point to point in a very good crystal using a novel mixed k-space and r-space probe of bulk heterogeneity combining high wave number resolution with micrometer-scale spatial resolution.

The scanning µXRD is similar to transmission electron microscopy (TEM), but without the complication of electron beam damage. In fact, the electron beam heating of the thin crystals can change the mobile oxygen content y and generate transient non-equilibrium surface structures that mask intrinsic bulk effects. Therefore information of finite size ordering properties as well as temperature dependent properties is not achievable by TEM.

The application of scanning µXRD for imaging the nanoscale phase separation in $YBa_2Cu_3O_{6.33}$ show the presence of a phase of nano-scale Ortho-II puddles embedded in an insulating oxygen poor background in a sample very close to the superconductor-insulator phase transition. The network of the Ortho-II puddles is made of different puddles ranging between 3 and 12 oxygen chains. The scanning µXRD and the temperature treatments show that the local structure in YBCO deviates from the ideal one. We have seen a relevant structure dependence with temperature in agreement with previous high resolution XRD data, where temperature treatments modify the density and elongation of the Cu-O chains [43-46]. The formation of nanoscale ordered grains in YBCO with Ortho-II lattice in the basal plane is highly relevant for the electronic structure of the system and for models of superconductor-to-insulator transition [19,20]. The oxygen chain puddles have the Ortho-II lattice superstructure, like the $YBa_2Cu_3O_{6.5}$ which shows band folding and Fermi surface reconstruction measured by photoemission and quantum oscillations experiments [48,49]. The granular structure of Ortho-II puddles favors the 7K superconductivity in fact in the rapidly quenched sample from T> 450K where the Ortho-II puddles are not formed the critical temperature drops to zero.

The nanoscopic phase separation of the oxygen puddles described here using µXRD, could explain why zero and transverse field µSR experiments have shown the coexistence of antiferromagnetic short range magnetism with superconductivity in YBCO with 6.37<y<6.39 [50]. In fact, since the average oxygen concentration is 6.33 and the local oxygen concentration in the Ortho-II puddles is 6.5, we deduce that in the intercalated spatial portions the oxygen concentration y has to be smaller than 6.33. In this way the oxygen puddles are expected to cover 66% of the space; in contrast with the not percolating nanoscopic magnetic portions. This explains why neutron scattering experiments do not find a static anti-ferromagnetism [50]. On the other hand, both neutron scattering and µSR experiments agree with the presence of two energy scales and short-ranged correlations. A similar granular disorder has been found also in single crystals of $La_2CuO_{4+y}$ [11,22,38,39] and of $Bi_2Sr_2CaCu_2O_{8+y}$ [25] using scanning µXRD.

Our non-invasive technique µXRD probes portions of the single crystals of one micron size, much smaller than those seen by neutron diffraction and previous X-ray diffraction [26] and therefore allows us to measure the formation of nanocrystalline puddles with unprecedented spatial resolution.

## 5. Conclusion

We have used scanning µXRD to probe the microscopic spatial inhomogeneity in $YBa_2Cu_3O_{6.33}$, which cannot be visualized through standard averaged X-ray diffraction. The spatial resolution of 1 micron used in our measurements allowed us to clearly visualize the granular nanoscopic phase separation due to the formation of the network of Ortho-II puddles made of oxygen ordered chains. Furthermore, this work shows that the size of the nanoscale oxygen ordered puddles in YBCO near the superconductor-insulator phase transition is much smaller than in $La_2CuO_{4+y}$ with a larger misfit-strain [7,8].

*Acknowledgments*
We thank Ruixing Liang, D.-A. Bonn, and Walter-N. Hardy of the Department of Physics of the University of British Columbia, for proving us with the crystals and for helpful discussions.

*References*

[1] A. Bianconi, *Solid State Communications* 89, 933 (1994)

[2] M. K. Wu, J. R. Ashburn, C. J. Torng, P. H. Hor, R. L. Meng, L. Gao, Z. J. Huang, Y. Q. Wang, and C. W. Chu, *Phys. Rev. Lett.* 58, 908 (1987).

[3] J. Zaanen and O. Gunnarsson, Physical Review B 40, 7391 (1989)

[4] A. Bianconi, A. Congiu-Castellano, M. De Santis, P. Rudolf, P. Lagarde, A. M. Flank, and A. Marcelli, *Solid State Communications* 63, 1009 (1987)

[5] T. Egami and S. J. L. Billinge, *Underneath the Bragg Peaks,* Volume 16: Structural Analysis of Complex Materials (Pergamon Materials Series) (Pergamon, 2003).

[6] A. Bianconi, N.L. Saini, A. Lanzara, M. Missori, T. Rossetti, H. Oyanagi, H. Yamaguchi, K. Oka, and T. Ito, *Physical Review Letters* 76, 3412 (1996).

[7] A. Bianconi, N.L Saini, S. Agrestini, D. Di Castro, G. Bianconi *International Journal of Modern Physics B* 14, 3342 (2000); A. Bianconi, D. Di Castro, G. Bianconi et al. *Physica C* 341, 1719 (2000)

[8] A. Bianconi, S. Agrestini, G. Bianconi, D. Di Castro, and N. L. Saini, *Journal of Alloys and Compounds* 317-318, 537 (2001)

[9] P. Littlewood, *Nature Materials* 10, 726 (2011).

[10] T. H. Geballe, & M. Marezio *Physica C: Superconductivity*, 469, 680 (2009).

[11] N. Poccia, A. Ricci, G. Campi, M. Fratini, A. Puri, D. D. D. Gioacchino, A. Marcelli, M. Reynolds, M. Burghammer, N. L. L. Saini, A. Bianconi, *Proc. Nat. Acad. Sci. USA* 109, 15685 (2012).

[12] L. P. Gor'kov & G. B. Teitel'baum *Phys. Rev. B*, 82, 020510 (2010).

[13] K. I. Kugel, A. L. Rakhmanov, A. O. Sboychakov, N. Poccia, and A. Bianconi *Phys. Rev. B*. 78, 165124 (2008)

[14] D. Innocenti, A. Ricci, N. Poccia, G. Campi, M. Fratini, and A. Bianconi, *J. of Superconductivity and Novel Magn*. 22, 529 (2009)

[15] A. Bianconi and M. Missori, *Solid State Communications* 91, 287 (1994).

[16] A. Perali, A. Bianconi, A. Lanzara, and N. L. Saini, *Solid State Communications* 100, 181 (1996)

[17] V Kresin, Y Ovchinnikov, & S Wolf *Physics Reports,* 431, 231 (2006)

[18] E. V. L. de Mello *EPL (Europhysics Lett.)* 98, 57008 (2012)

[19] G. Bianconi *Phys. Rev. E*, 85, 061113 (2012).

[20] G. Bianconi *Journal of Statistical Mechanics: Theory and Experiment* 2012, P07021 (2012).

[21] J. Zaanen *Nature* 466, 825 (2010)

[22] M. Fratini, N. Poccia, A. Ricci, G. Campi, M. Burghammer, G. Aeppli, and A. Bianconi, *Nature* 466, 841 (2010).

[23] N. L. Saini, H. Oyanagi, T. Ito, V. Scagnoli, M. Filippi, S. Agrestini, G. Campi, K. Oka, and A. Bianconi, *The European Physical Journal B - Condensed Matter and Complex Systems* 36, 75 (2003)

[24] E. S. Božin, G H Kwei, H Takagi, & S J L Billinge *Physical Review Letters,* 84, 5856 (2000).

[25] N. Poccia, G. Campi, M. Fratini, A. Ricci, N. L. Saini, and A. Bianconi *Phys. Rev. B* 84, 100504 (2011)

[26] G. Campi, E. Cappelluti, T. Proffen, X. Qiu, E. S. Bozin, Billinge, S. Agrestini, N. L. Saini, and A. Bianconi, *The European Physical Journal B - Condensed Matter and Complex Systems* 52, 15 (2006).

[27] V. Ksenofontov, G. Wortmann, S.A. Medvedev, V. Tsurkan, J. Deisenhofer, A. Loidl, and C. Felser, *Phys. Rev. B* 84, 180508 (2011)

[28] A. Charnukha, J. Deisenhofer, D. Pröpper, M. Schmidt, Z. Wang, Y. Goncharov, A. N. Yaresko, V. Tsurkan, B. Keimer, A. Loidl, et al., *Phys. Rev. B* 85, 100504 (2012).

[29] A. Ricci, N. Poccia, G. Campi, B. Joseph, G. Arrighetti, L. Barba, M. Reynolds, M. Burghammer, H. Takeya, Y. Mizuguchi, et al., *Phys. Rev. B* 84, 060511 (2011)

[30] E Dagotto *Nanoscale phase separation and colossal magnetoresistance : the physics of manganites and related compounds.* Springer (2003).

[31] J. D. Jorgensen, S. Pei, P. Lightfoor, H. Shi, A. P. Paulikas, and B. W. Veal, P*hysica C: Superconductivity* 167, 571 (1990)

[32] B. W. Veal, A. P. Paulikas, H. You, H. Shi, Y. Fang, and J. W. Downey, *Phys. Rev. B* 42, 6305 (1990).

[33] G. Ceder, R. McCormack, and D. de Fontaine, Physical Review B 44, 2377 (1991)

[34] H. Friis Poulsen, N. Hessel Andersen, J. Vrtting Andersen, H. Bohrt, and O. G. Mouritsen, *Nature* 349, 594 (1991).

[35] G. Yu, C. H. Lee, A. J. Heeger, N. Herron, E. M. McCarron, L. Cong, G. C. Spalding, C. A. Nordman, and A. M. Goldman, *Phys. Rev. B* 45, 4964 (1992).

[36] S. Sanna, G. Allodi, G. Concas, A. D. Hillier, and R. De Renzi, *Phys. Rev. Lett.* 93, 207001 (2004)

[37] C. L. Johnson, J. K. Bording, and Y. Zhu, *Phys. Rev. B* 78, 014517 (2008)

[38] N. Poccia, M. Fratini, A. Ricci, G. Campi, L. Barba, A. Vittorini-Orgeas, G. Bianconi, G. Aeppli, and A. Bianconi, *Nature Materials* 10, 733 (2011).

[39] N. Poccia, A. Bianconi, G. Campi, M. Fratini, and A. Ricci, *Superconductor Science and Technology* 25, 124004+ (2012).

[40] Y. Nakazawa, M. Ishikawa, T., Takabatake, K.-i. Koga, & K. Terakura, *Japanese Journal of Applied Physics* 26, L796 (1987).

[41] R. Liang, D. A. Bonn, and W. N. Hardy, *Physica C,* 336, 57 (2000); R. Liang et al., *Physica C,* 383, 1 (2002).

[42] J. Strempfer, I. Zegkinoglou, U. Rütt, Zimmermann, C. Bernhard, C. T. Lin, Th, and B. Keimer, *Physical Review Letters* 93, 157007 (2004)

[43] M. von Zimmermann, J. R. Schneider, T. Frello, N. H. Andersen, J. Madsen, M. Käll, H. F. Poulsen, R. Liang, P. Dosanjh, W. N. Hardy, *Phys. Rev. B* 68, 104515 (2003)

[44] Ruixing Liang, D.A. Bonn, W.N. Hardy *Phil. Mag.* 92, 2356 (2012).

[45] Ruixing Liang, D.A. Bonn, W.N. Hardy, *Physica C*. 304, 105 (1998).

[46] Ruixing Liang, D.A. Bonn, W.N. Hardy, J.C. Wynn, K.A. Moler, L. Lu, S. Larochelle, L. Zhou, M. Greven, L. Lurio, S.G.J. Mochrie *Physica C* 383, 1 (2002).

[47] J. Garcia, A. Bianconi, M. Benfatto, and C. R. Natoli, Le *Journal de Physique Colloques* 47, C8 (1986)

[48] Y. Sassa, M. Radović, M. Maansson, E. Razzoli, X. Y. Cui, S. Pailhès, S. Guerrero, M. Shi, P. R. Willmott, F. M. Granozio, et al., *Phys. Rev. B* 83, 140511 (2011)

[49] D. Fournier, G. Levy, Y. Pennec, J. L. McChesney, A. Bostwick, E. Rotenberg, R. Liang, W. N. Hardy, D. A. Bonn, I. S. Elfimov, et al., *Nature Physics* 6, 905 (2010)

[50] C. Stock, W. J. L. Buyers, Z. Yamani, C. L. Broholm, J. H. Chung, Z. Tun, R. Liang, D. Bonn, W. N. Hardy, and R. J. Birgeneau, *Phys. Rev. B* 73, 100504 (2006).